# Examining Experimental Raman Mode Behavior in Mono- and Bilayer 2H-TaSe$_2$ via Density Functional Theory

Sugata Chowdhury[†,‡], Heather M. Hill[†], Albert F. Rigosi[†], Andrew Briggs[†], Helmuth Berger[§], David B. Newell[†], Angela R. Hight Walker[†], and Francesca Tavazza[†]*

[†]National Institute of Standards and Technology (NIST), Gaithersburg, MD 20899, United States

[‡]Department of Physics and Astronomy, Howard University, Washington, DC 20059, United States

[§]École Polytechnique Fédérale de Lausanne (EPFL), Institut de Physique des Nanostructures, CH-1015 Lausanne, Switzerland

*Keywords: charge density waves, density functional theory, transition metal dichalcogenides, Raman spectroscopy, van der Waals interactions.*

**ABSTRACT:** Tantalum diselenide (TaSe$_2$) is a metallic transition metal dichalcogenide whose equilibrium structure and vibrational behavior strongly depends on temperature and thickness, including the emergence of charge density wave (CDW) states at very low T. In this work, observed modes for mono- and bi-layer are described across several spectral regions and compared to the bulk ones. Such modes, including an experimentally observed forbidden Raman mode and low frequency CDW modes, are then matched to corresponding density functional theory (DFT) predicted vibrations, to unveil their inner working. The excellent match between experimental and computational results justifies the presented vibrational visualizations of these modes. Additional support is provided by experimental phonons seen in Raman spectra as a function of temperature and thickness. These results highlight the importance of understanding interlayer interactions and their effects on mode behaviors.

Layered two-dimensional (2D) materials are the subject of many research pursuits for their versatile and novel properties [1-4]. Among those 2D materials, TaSe$_2$ attracts interest due to its ability to host charge density wave (CDW) states, yielding potential functions in quantum information science [5, 6]. Applications for quantum devices often require fabrication of atomically thin material to maximize the tunability of the material's properties [7]. For layered materials, both van der Waals (vdW) interactions and quantum confinement make significant contributions to the layer-dependent properties. It then follows that a comparison between monolayer (1L), bilayer (2L), and bulk systems should be made to provide a general understanding of what modes emerge and how their physical behavior differs with thickness.

Generally, layered 2H-TaSe$_2$ exhibits an incommensurate CDW (IC-CDW) phase between 122 K and 90 K, as well as a commensurate CDW (C-CDW) phase below 90 K [8, 9]. Also, since the superconducting phase transition of 2H-TaSe$_2$ occurs at 0.2 K, the corresponding phase does not coexist with CDW phase, rendering this material an ideal system for studying the layer-dependence of such quantum phase transitions. To date, several studies have focused on the evolution of electronic structures and CDW formation [10-13], but few explore low-frequency modes [14]. Though various modes in atomically thin 2H-TaSe$_2$, as well as others similar materials like NbSe$_2$, are described to some degree by Raman spectra in other work [15-20], many modes elude sufficient visualization, warranting an investigation like the one presented here [21-28].

In this work, experimental data from Raman spectroscopy and calculated predictions from density functional theory (DFT) were used to examine observed low-frequency CDW modes and the forbidden mode in layered 2H-TaSe$_2$. Specifically, DFT was employed to calculate the appropriate electronic structures of 1L, 2L, and bulk material until a suitable agreement was found for well-known, higher frequency modes (namely, the $E_{2g}^1$ and $A_{1g}$ modes). This agreement served as a support that subsequent modes found in the mid- and low-frequency ranges could be accurately described by the same DFT model. The various vibrational modes are discussed for both 300 K and temperatures below the CDW phase transitions, with discussions around the temperature-dependent behavior of those modes. Significant differences between the phonons of atomically thin and bulk systems were observed, and these modes are described visually since such descriptions are lacking in the literature.

Calculations were obtained with Quantum ESPRESSO, with the core plane wave functions provided by the Plane-

Wave Self-Consistent Field (PWscf) component [29-31]. Perdew-Zunger (PZ) exchange and correlation functionals were used for the phonon calculations within the applied local-density approximations (LDAs) [32]. Applying LDAs yielded a more accurate description of the optical properties of TaSe$_2$ than the general-gradient approximation (GGA), despite the general tendency of LDAs to underestimate the exchange energy and overestimate the correlation energy [33]. Norm-conserving pseudopotentials were used for describing the interactions between the core and valence electrons [34, 35].

All DFT calculations were done at a thermal temperature of 0 K, but to model the temperature-dependent formation of various Raman modes, the *electronic* temperature was modulated by tuning the smearing factor σ, a parameter which describes the Fermi-Dirac distribution. The effect of temperature on the phonon properties was also explored [36-41]. The approach of modeling real temperature effects with electronic temperature variations was validated by computing the lattice expansion as a function of temperature and comparing it with experimental results.

Ground state calculations were performed by relaxing atomic positions and lattice vectors of 1L 2H-TaSe$_2$ (point group $D_{3h}$). The optimized lattice constants ($a_{TaSe2}$ = 0.337 nm and $c_{TaSe2}$ = 1.23 nm) are within reasonable ranges when compared with other computational studies [42, 43]. The 1L supercell was constructed having 9 unit cells (3 × 3 × 1) and used a sufficiently large vacuum (20 Å) in the vertical direction to negate any interaction between neighboring supercells. The kinetic energy cutoff of the plane-wave expansion was 520 eV. All geometric structures were fully relaxed until the force on each atom was less than 0.002 eV/Å and the energy-convergence criterion was 10$^{-7}$ eV. For the unit cell and supercell structure relaxation, a 16 × 16 × 16 and 3 × 3 × 8 $k$-point grid was used, respectively. For phonon calculations, a 4 × 4 × 4 uniform $q$-grid was used for the unit cell and only performed gamma point phonon calculations for the supercell. Further details are included in the Supporting Information (SI).

For the experimental data, mechanically exfoliated, single crystals of home-grown 2H-TaSe$_2$ on Si/SiO$_2$ substrates (300 nm oxide layer) were prepared using adhesive tape to peel layers off. During Raman spectrum acquisition, a 515 nm laser excitation was used at sample temperatures between 5 K and 300 K. Both 1L and 2L samples were identified with optical contrast and well-known Raman modes ($E_{2g}^1$ and $A_{1g}$). The scattered light was collected through a triple-grating spectrometer to enable low-frequency measurements (with detection as low as 10 cm$^{-1}$). Modes in Raman spectra were characterized by Lorentzian profiles to extract position and width information, unless otherwise stated.

A comparison of DFT predictions to experimental Raman spectra is shown in Fig. 1, where the predictions were obtained from the calculated structures for the supercell (3 × 3 × 1) of 2H-TaSe$_2$, for the 1L, 2L, and bulk cases at the endpoint temperatures 10 K and 300 K and compared to Raman spectra obtained at 5 K and 300 K. Although bulk 2H-TaSe$_2$ is predicted to exhibit twelve phonon modes, represented by Γ ($D_{6h}$) = A$_{1g}$ + 2A$_{2u}$ + B$_{1u}$ + 2B$_{2g}$ + E$_{1g}$ + 2E$_{1u}$ + E$_{2u}$ + 2E$_{2g}$, only four are Raman active: A$_{1g}$, E$_{1g}$, and two E$_{2g}$ modes.

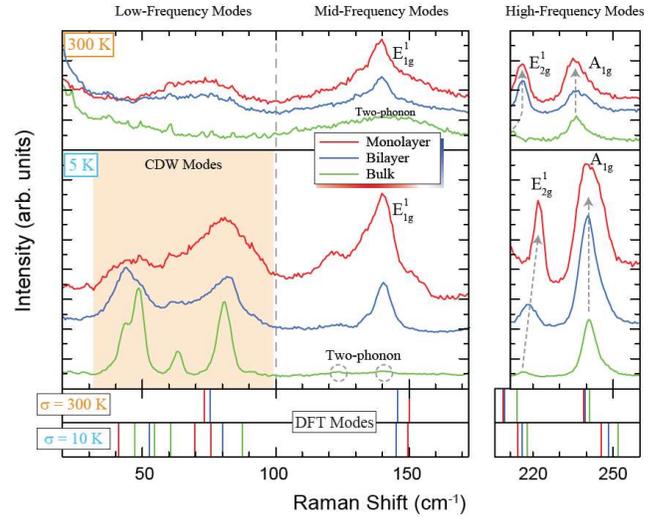

Figure 1. Experimental Raman spectra are presented here for the 1L (red curves), 2L (blue curves), and bulk (green curves) TaSe$_2$ at two temperatures: 300 K (top panel) and 5 K (lower panel). Two vertically shortened panels are displayed at the bottom for DFT modes at the two temperatures with corresponding color schemes. The comparisons of these spectra are subdivided into three intervals – high-, mid-, and low-frequency region. The shaded orange region in the 5 K measurements highlight CDW modes.

The DFT-calculated peak positions, shown in the lowest panel of Fig. 1, align well with the experimental data, providing support for this model in terms of characterizing other lesser-known modes. There are several significant changes when transitioning from room temperature (300 K) to below the CDW phase transitions (5 K) that will be detailed in due course. Raman spectra will be evaluated within each of the following three spectral regions: (a) high frequency range (above 200 cm$^{-1}$), (b) mid-frequency range (100 cm$^{-1}$ to 175 cm$^{-1}$), and (c) low frequency range (below 100 cm$^{-1}$). The exact spectral region bounds were chosen arbitrarily for the sake of orderliness, but the high-frequency range was specifically selected to validate the DFT model used for subsequent descriptions of modes found in the other two spectral regions.

Using the experimental data in Fig. 2 (a) as a reference, the evolution of the theoretically predicted high frequency modes $A_{1g}$ and $E_{2g}^1$ in the 1L case is shown in Fig. 2 (b). Both 1L modes display a blueshift with decreasing temperature, similar to their bulk counterparts, though the nature of those modes is inherently different between 1L and bulk, as discussed below. The experimental Raman spectra and DFT results agree within 10 cm$^{-1}$ for the $E_{2g}^1$ mode, and within 2 cm⁻¹ for the $A_{1g}$ mode, which exhibited a slightly smaller blueshift than the $E_{2g}^1$ mode. The differences in the trend for $A_{1g}$ could be due to the experimental data being taken on a sample mounted on a substrate. This consideration would be consistent with $A_{1g}$ being an out-of-plane mode, thus being strongly influenced by its surrounding environment.

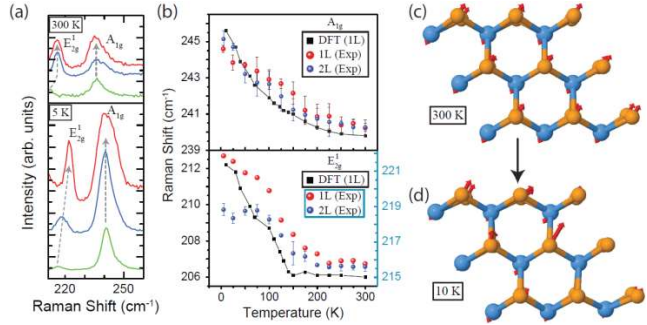

Figure 2. (a) The two experimentally observed high-frequency modes for TaSe₂ are shown for 1L, 2L, and bulk. (b) In the top panel, the theoretically derived data of the $A_{1g}$ mode is shown as a function of electronic temperature (σ) and compared with experimental data (red and blue circles for 1L and 2L, respectively). In the bottom panel, the theoretically derived data of the $E_{2g}^1$ mode is observed to abruptly begin shifting as the temperature decreases (represented by the left vertical axis). Corresponding experimental data are shown with the same data point format and represented by the right vertical axis. The error bars indicate a 1σ standard deviation from the determined peak value using Lorentzian fits, and in the bottom panel, are smaller than some of the data points. (c) Illustrations of the 1L Ta atoms' ($E_{2g}^1$) vibration direction are shown at 300 K and (d) 10 K. Note that despite the appearance of this mode as simply displaying a blueshift with decreasing temperature, its inherent vibrations at the two temperatures are different.

Another experimental observation was made in the temperature-dependent change of frequency exhibited by the $A_{1g}$ and $E_{2g}^1$ modes. In the 1L case, the two peaks blueshifted with decreasing temperature by 4.1 cm⁻¹ ± 1.2 cm⁻¹ and 6.2 cm⁻¹ ± 0.2 cm⁻¹, respectively, where the error indicates a 1σ standard deviation. For the bulk case, those two shifts were 5.0 cm⁻¹ ± 0.2 cm⁻¹ and 6.7 cm⁻¹ ± 0.2 cm⁻¹, respectively. The different behavior between the 1L and bulk temperature-dependent shifts can be explained consistently with results reported in other work [28]. To summarize, the $A_{1g}$ mode in 1L (and 2L) experiences a decreased force constant due to a weakening vdW interaction, which is why the temperature-dependent shifting of this mode is nearly identical for 1L, 2L, and bulk (to within experimental error). Second, the temperature-dependent shift of the $E_{2g}^1$ mode is greater in bulk compared to the 1L case, highlighting the decreasing strength of long-range Coulomb interactions as the material became thinner [44-46].

The most significant temperature-dependent similarities between the atomically thin material and bulk are the smooth behavior seen in the $A_{1g}$ mode and the abrupt onset behavior observed for the $E_{2g}^1$ mode. The slope discontinuity seen for the theoretically derived $E_{2g}^1$ mode in 1L (Fig. 2 (b)), as well as that seen theoretically and experimentally in bulk [28], is due to the Ta atoms abruptly changing vibration direction as they approach the CDW phase transition temperature. In Fig. 2 (c) and (d), the 1L $E_{2g}^1$ mode at 300 K and 10 K, respectively, exhibits a temperature-dependent frequency. For the two temperatures listed, the two corresponding frequencies characterizing the 1L $E_{2g}^1$ mode are in close spectral proximity, but both frequencies are actually represented by different Ta atom vibration directions. In the 1L case, vdW interactions do not have as strong a contribution to the creation of a coherent vibration direction, whereas in the bulk case, all Ta atoms abruptly rotate their vibration direction by 30° as they cross below the critical temperature (122 K). Additional details about the changes in the $E_{2g}^1$ mode during the CDW phase transition may be found in the SI. Overall, the DFT model can be validated by its agreement with the trends seen in the experimental data.

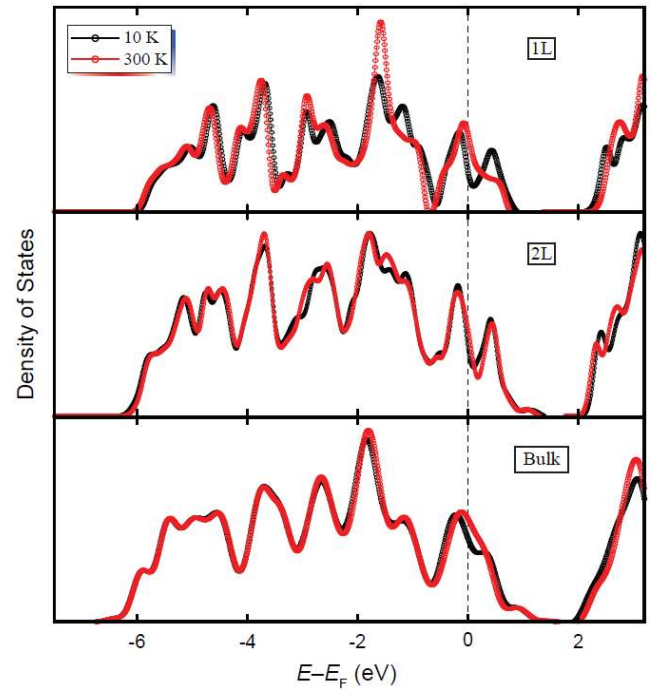

Figure 3. The total density of states (DOS) for the 1L, 2L, and bulk systems is shown in a top, middle, and bottom panel, respectively. Furthermore, the DOS at 300 K and 10 K are calculated and plotted in red and black curves, respectively. From the comparison between the 1L and bulk DOS, one may see significant changes at the Fermi level, with significant contributions coming from the d-orbital of the Ta atoms (refer to the SI).

To obtain a clearer understanding of the DFT model used, the total density of states (DOS) was calculated. In

Fig. 3, the DOS for the 1L, 2L, and bulk systems are shown. From the comparison between the atomically thin and bulk system DOS, one may conclude that there are significant changes at the Fermi level and that significant contributions originate from the *d*-orbital of the Ta atoms (see SI). The Ta atoms' contribution strongly depends on the position of the atoms. Changes at the Fermi level can be attributed to the edge atoms of the triangle sublattice, whereas central Ta atoms contribute most strongly in the far lower energies of the valance band [47]. Surprisingly, for the bulk crystal, central Ta atoms in the striped formations do not contribute to the DOS [28]. Additionally, both the valence and conduction bands undergo a more significant change in the atomically thin cases compared with the bulk counterpart.

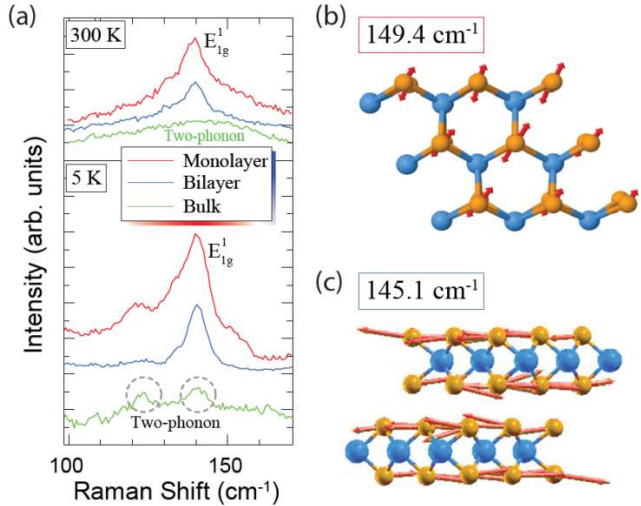

Figure 4. (a) Experimentally observed Raman spectra of 1L, 2L, and bulk for the mid-frequency range. (b) The forbidden mode for 1L is visualized here and compared with (c) the side-view visualization of the 2L forbidden mode. Both visualizations appear very similar despite the slight redshift in both the DFT and experimental data with an added layer.

The same DFT model is used to gain a better understanding for the mid-frequency range of the Raman spectra, where experimental data show the emergence of a forbidden mode ($E_{1g}^1$) present in the 1L and 2L Raman spectra but not in the bulk spectrum (Fig. 4 (a)). At room temperature, there appears to be an overlapping of the $E_{1g}^1$ and two-phonon modes (dominant in the bulk but not in the 1L spectrum), whereas at 5 K, the overlap becomes less apparent, with the $E_{1g}^1$ mode becoming more accentuated and the two-phonon mode in bulk becoming more distinct [48-51]. The position and intensity of the forbidden $E_{1g}^1$ mode in the 1L and 2L cases were found to be temperature-independent (see SI). The DFT-calculated position of the $E_{1g}^1$ mode is 149.4 cm$^{-1}$ (at 10 K) and this agrees well with previously reported experimental 1L work [44]. The evolution of the temperature-dependent two-phonon mode for bulk 2H-TaSe$_2$ systems has been dis-cussed in a recent publication [28]. One explanation for the observation of a forbidden $E_{1g}^1$ mode in the 2D limit is the introduction of a broken symmetry in the 1L case (lack of a periodic third dimension) [48]. For the case of 2L, the same argument holds, though to a slightly lesser extent. The broken symmetry still exists with a second layer, though the intensity may not be as high. As shown in Fig. 4 (b) and (c), the visualization of the forbidden mode does not appear to change with the addition of a second layer, despite the slight redshift in both DFT and experimental data.

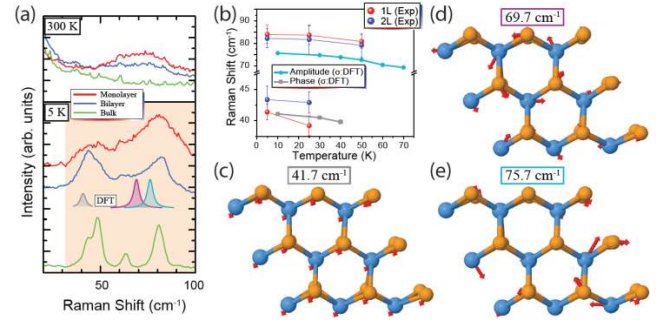

Figure 5. (a) The experimental Raman spectra from the low-frequency region are shown vertically translated. (b) Two modes in the 1L case are calculated as a function of electronic temperature (σ) and determined to be a phase and amplitude mode. The DFT predictions are compared with 1L and 2L extracted mode positions, with errors bars indicating a 1σ standard deviation from the Lorentzian fits. (c)-(e) Corresponding illustrations of the 1L low-frequency modes calculated with DFT are shown in order of increasing frequency, determined at a set electronic temperature of 10 K.

Lastly, results for the low-frequency region are examined. In Figure 5 (a), the experimental 1L and 2L spectrum is superposed with three DFT-calculated modes most likely to describe the observed behavior. The lowest frequency mode is predicted at 41.7 cm$^{-1}$ and attributed to a phase mode, which exhibits a translational behavior with all atomic layers vibrating in the same direction [28]. The mode emerges at 40 K in the C-CDW phase and blueshifts as the temperature decreases as shown in Fig. 5 (b). Experimental Raman data were taken at 25 K, 50 K, 75 K, and all incremental temperatures until 300 K (in steps of 25 K). Data at 50 K for both the 1L and 2L systems does not show this phase mode, indicating that the emergence temperature is between 25 K and 50 K. An illustration of the phase mode is shown in Fig. 5 (c). The second mode, seen in Fig. 5 (d), is a circular mode at 69.7 cm$^{-1}$, and it is independent of temperature and atomic displacement. The third and final mode predicted in 1L TaSe$_2$, seen in Fig. 5 (e), emerges near 70 K and blueshifts with decreasing temperature and can be described as an amplitude mode, taking on a more distorted form of circular vibration in the opposite direction to the mode at 69.7 cm$^{-1}$ [28]. The predicted temperature-dependent behavior of this third mode in 1L is shown in light blue, with the ex-perimental data for both 1L and 2L showing a similar trend. The Raman spectra taken at 75 K did not appear to include this mode, falling in line with the DFT prediction.

In the bulk case, four CDW modes emerge – two phase and two amplitude modes with larger intensities than the Raman $A_{1g}$ and $E_{2g}^1$ modes [28]. Though the 1L and 2L modes in this region have intensities similar to their bulk counterparts, it is important to note that the bulk modes are completely different lattice modes than those in the 1L or 2L spectra. For all four modes in the bulk case, the Se atoms are vibrating in opposite directions to each other and the Ta atoms do not vibrate at all. However, in the atomically thin case, all atoms are vibrating in some ordered fashion [52, 53]. These predicted differences between atomically thin and bulk vibrations is another example of how increased vdW interactions in bulk systems (i.e. increased interactions with neighboring layers) may restrict atoms from oscillating as strongly as those found in 1L or 2L systems.

In this work, observed low-frequency CDW modes and the forbidden mode in layered 2H-TaSe$_2$ were examined via Raman spectroscopy and DFT. Experimental data from the more well-known, high-frequency modes ($E_{2g}^1$ and $A_{1g}$ modes) in 1L, 2L, and bulk TaSe$_2$ were compared with DFT predictions to validate a suitable model for exploring modes not fully elucidated in the literature. Consequently, modes found in the mid- and low-frequency ranges were accurately described by the same DFT model, with the various vibrational modes discussed for temperatures between 300 K and 5 K. The significant differences observed between the modes of atomically thin and bulk systems were subsequently rendered visually. Overall, the methods underlying this work can be extended to describe and visualize modes in other 2D materials while also highlighting the effects that vdW interactions have on the optical properties of those materials systems.

## ASSOCIATED CONTENT

**Supporting Information**. Crystal structure of bulk and monolayer 2H-TaSe2, Structural parameters, Temperature-dependent evolution, and PDOS for orbitals provided as Supporting Information. This material is available free of charge via the Internet at http://pubs.acs.org.

## AUTHOR INFORMATION

### Corresponding Author
* francesca.tavazza@nist.gov

### Author Contributions
S.C. performed all theoretical calculations. H.M.H and A.F.R. performed experiments and prepared samples. A.B. and H.B. provided the material. D.B.N., A.R.H.W., and F.T. assisted with the analyses, support, and general project oversight. The manuscript was written through contributions of all authors. The manuscript was written through contributions of all authors.

### Funding Sources
NSF grant DMR-180119

### Notes
Commercial equipment, instruments, and materials are identified in this paper in order to specify the experimental procedure adequately. Such identification is not intended to imply recommendation or endorsement by the National Institute of Standards and Technology or the United States government, nor is it intended to imply that the materials or equipment identified are necessarily the best available for the purpose. The authors declare no competing interests.

## ACKNOWLEDGMENT
The authors would like to thank S. Payagala and A. Biacchi for their assistance in the NIST internal review process. Support was provided by the National Science Foundation through grant DMR-180119. Work presented herein was performed, for a subset of the authors, as part of their official duties for the United States government. Funding is hence appropriated by the United States Congress directly.

Insert Table of Contents artwork here

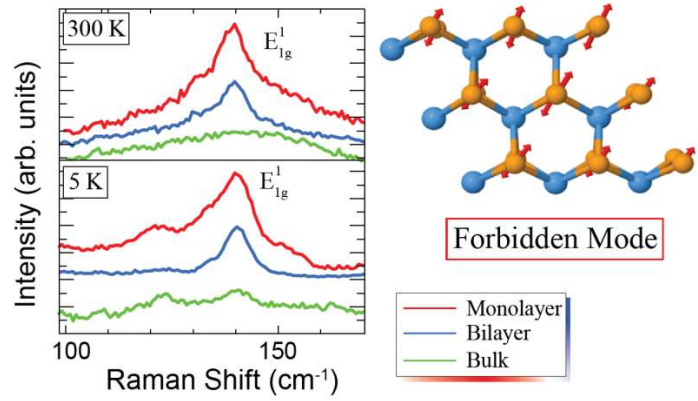